\documentclass[aps,twocolumn]{revtex4}
\usepackage{bm}
\usepackage{graphicx}
\usepackage{epsfig}
\begin{document}

\title{Photo-induced pure spin currents in quantum wells}
\author{S.A.~Tarasenko}
\author{E.L.~Ivchenko}
\affiliation{A.F.~Ioffe Physico-Technical Institute, 194021
St.~Petersburg, Russia }
\begin{abstract}
As is well known the absorption of circularly polarized light in
semiconductors results in optical orientation of electron spins
and helicity-dependent electric photocurrent, and the absorption
of linearly polarized light is accompanied by optical alignment of
electron momenta. Here we show that the absorption of unpolarized
light in a quantum well (QW) leads to generation of a pure spin
current, although both the average electron spin and electric
current are vanishing.
\end{abstract}
\maketitle

\section{Introduction}

Spin and charge are among the basic properties of elementary
particles such as an electron, positron and proton. The
perturbation of a system of electrons by an electric field or
light may lead to a flow of the particles. The typical example is
an electric current that represents the directed flow of charge
carriers. Usually the electric currents do not entail a
considerable spin transfer because of the random orientation of
electron spins. However, the charge current can be accompanied by
a spin current as it happens, e.g., under injection of
spin-polarized carriers from magnetic materials or in the
optical-orientation-induced circular photogalvanic effect.
Furthermore, there exists a possibility to create a pure spin
current which is not accompanied by a net charge transfer. This
state represents a non-equilibrium distribution when electrons
with the spin ``up'' propagate mainly in one direction and those
with the spin ``down'' propagate in the opposite direction. In
terms of the kinetic theory, it can be illustrated by a spin
density matrix with two nonzero components, $\rho(s, \bm{k}; s,
\bm{k})= \rho(\bar{s}, - \bm{k}; \bar{s}, - \bm{k})$, where $s$
and $\bm{k}$ are the electron spin index and the wave vector, and
$\bar{s}$ means the spin opposite to $s$. Spin currents in
semiconductors can be driven by an electric field acting on
unpolarized free carriers which undergo a spin-dependent
scattering. This is the so-called spin Hall effect where a pure
spin current appears in the direction perpendicular to the
electric field. The spin currents can be induced as well by
optical means as a result of interference of one- and two-photon
coherent excitation with a two-color electro-magnetic field or
under interband optical transitions in non-centrosymmetrical
semiconductors~[1].

Here we show that pure spin currents, accompanied neither by
charge transfer nor by spin orientation, can be achi\-ev\-ed under
absorption of linearly polarized or unpolarized light in
semiconductor low-dimensional systems. Particularly, the effect is
considered for interband and free-carrier optical transitions in
semiconductor QWs.

In general, the flux of electron spins can be characterized by a
pseudo-tensor $\hat{\bm{F}}$ with the components
$F_{\beta}^{\alpha}$ describing the flow in the $\beta$ direction
of spins oriented along $\alpha$, with $\alpha$ and $\beta$ being
the Cartesian coordinates. In terms of the kinetic theory such a
component of the spin current is contributed by a non-equilibrium
correction $\propto \sigma_{\alpha} k_{\beta}$ to the electron
spin density matrix, where $\sigma_{\alpha}$ is the Pauli matrix.
We demonstrate that the appearance of a pure spin current under
optical pumping is linked with two fundamental properties of
semiconductor QWs, namely, the {\it spin splitting} of energy
spectrum linear in the wave vector and the spin-sensitive {\it
selection rules} for optical transitions.

\section{Interband optical transitions}

\begin{figure}[b]
\leavevmode \epsfxsize=0.6\linewidth
\centering{\epsfbox{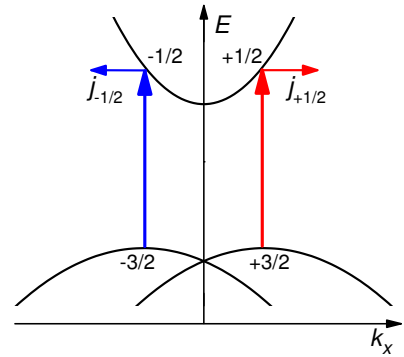}} \caption{Microscopic origin
of pure spin current induced by interband photoexcitation.}
\end{figure}

The effect is most easily conceivable for direct transitions
between the heavy-hole valence subband $hh1$ and conduction
subband $e1$ in QWs of the C$_s$ point symmetry, e.g. in (113)- or
(110)-grown QWs. In such structures the spin component along the
QW normal $z$ is coupled with the in-plane electron wave vector.
This leads to $\bm{k}$-linear spin-orbit splitting of the energy
spectrum as sketched in Fig.~1, where heavy hole subband $hh1$ is
split into two spin branches $\pm 3/2$. As a result they are
shifted relative to each other in the $\bm{k}$ space. In the
reduced-symmetry structures, the spin splitting of the conduction
subband  is usually smaller than that of the valence band and not
shown in Fig.~1 for simplicity. Due to the selection rules the
direct optical transitions from the valence subband $hh1$ to the
conduction subband $e1$ can occur only if the electron angular
momentum changes by $\pm 1$. It follows then that the allowed
transitions are $|+3/2 \rangle \rightarrow |+1/2 \rangle$ and
$|-3/2 \rangle \rightarrow |-1/2 \rangle$, as illustrated in
Fig.~1 by vertical lines. Under excitation with linearly polarized
or unpolarized light the rates of both transitions are equal. In
the presence of spin splitting, the optical transitions induced by
photons of the fixed energy $\hbar\omega$ occur in the opposite
points of the $\bm{k}$-space for the spin branches $\pm 1/2$. The
asymmetry of photoexcitation results in a flow of electrons within
each spin branch. The corresponding fluxes $\bm{j}_{1/2}$ and
$\bm{j}_{- 1/2}$ are of equal strengths but directed in the
opposite directions. Thus, this non-equilibrium electron
distribution is characterized by the nonzero spin current
$\bm{j}_{\rm spin}$ = $(1/2) (\bm{j}_{1/2} - \bm{j}_{- 1/2})$ but
a vanishing charge current, $e (\bm{j}_{1/2} + \bm{j}_{- 1/2})=0$.

The direction $\beta$ of the photo-induced spin current and the
orientation $\alpha$ of transmitted spins are determined by the
form of spin-orbit interaction. The latter is governed by the QW
symmetry and can be varied. For QWs based on zinc-blende-lattice
semiconductors and grown along the crystallographic direction
$[110] \parallel z$, the light absorption leads to a flow along $x
\parallel [1\bar{1}0]$ of spins oriented along $z$. This component
of the electron spin flow can be estimated as
\begin{equation}\label{j_inter}
F_{x}^{z} = \gamma_{zx}^{(hh1)} \frac{\tau_e}{2 \hbar}
\frac{m_h}{m_e+m_h} \frac{\eta_{cv}}{\hbar\omega} I \:,
\end{equation}
where $\gamma_{zx}^{(hh1)}$ is a constant describing the
$\bm{k}$-linear spin-orbit splitting of the $hh1$ subband,
$\tau_e$ is the relaxation time of the spin current, $m_e$ and
$m_h$ are the electron and hole effective masses in the QW plane,
respectively, $\eta_{cv}$ is the light absorbance, and $I$ is the
light intensity. Note that the time $\tau_e$ can differ from the
conventional momentum relaxation time that governs the electron
mobility.

In (001)-grown QWs the absorption of linearly- or unpolarized
light results in a flow of electron spins oriented in the QW
plane. In contrast to the low-symmetry QWs considered above, in
(001)-QWs the $\bm{k}$-linear spin splitting of the valence
subband $hh1$ is small and here, for the sake of simplicity, we
assume the parabolic spin-independent dispersion in the $hh1$
valence subband and take into account the spin-dependent
contribution $\gamma_{\alpha \beta}^{(e1)} \, \sigma_{\alpha}
k_{\beta}$ to the electron effective Hamiltonian. Then, to the
first order in the spin-orbit coupling, the components of the pure
spin current generated in the subband $e1$ are derived to be
\begin{equation} \label{intband}
F^{\alpha}_{\beta} = \gamma_{\alpha\beta}^{(e1)}
\frac{\tau_e}{2\hbar} \frac{m_e}{m_e+m_h} \frac{\eta_{cv}}{\hbar
\omega} I \:.
\end{equation}

\section{Free-carrier absorption}

Light absorption by free carriers, or the Drude-like absorption,
occurs in doped semiconductor structures when the photon energy
$\hbar\omega$ is smaller than the band gap as well as the
intersubband spacing. Because of the energy and momentum
conservation the free-carrier optical transitions become possible
if they are accompanied by electron scattering by acoustic or
optical phonons, static defects etc. Scattering-assisted
photoexcitation with unpolarized light also gives rise to a pure
spin current. However, in contrast to the direct transitions
considered above, the spin splitting of the energy spectrum leads
to no essential contribution to the spin current induced by
free-carrier absorption. The more important contribution comes
from asymmetry of the electron spin-conserving scattering. In
semiconductor QWs the matrix element $V$ of electron scattering by
static defects or phonons has, in addition to the main
contribution $V_0$, an asymmetric spin-dependent term~[2]
\begin{equation}\label{V_asym}
V = V_0 + \sum_{\alpha\beta}V_{\alpha\beta} \,\sigma_{\alpha}
(k_{\beta} + k'_{\beta}) \:,
\end{equation}
where $\bm{k}$ and $\bm{k}'$ are the electron initial and
scattered wave vectors, respectively. Microscopically this
contribution is caused by the structural and bulk inversion
asymmetry similar to the Rashba/Dresselhaus spin splitting of the
electron subbands. The asymmetry of the electron-phonon
interaction results in non-equal rates of indirect optical
transitions for opposite wave vectors in each spin subband. This
is illustrated in Fig.~2, where the free-carrier absorption is
shown as a combined two-stage process involving electron-photon
interaction (vertical solid lines) and electron scattering
(dash\-ed horizontal lines). The scattering asymmetry is shown by
thick and thin dashed lines: electrons with the spin $+1/2$ are
preferably scattered into the states with $k_x > 0$, while
particles with the spin $-1/2$ are scattered predominantly into
the states with $k_x < 0$. The asymmetry causes an imbalance in
the distribution of photoexcited carriers in each subband $s=\pm
1/2$ over the positive and negative $k_x$ states and yields
oppositely directed electron flows $\bm{j}_{\pm1/2}$ shown by
horizontal arrows. Similarly to the interband excitation
considered in the previous section, this non-equilibrium
distribution is characterized by a pure spin current without
charge transfer.

\begin{figure}[t]
\leavevmode \epsfxsize=0.6\linewidth
\centering{\epsfbox{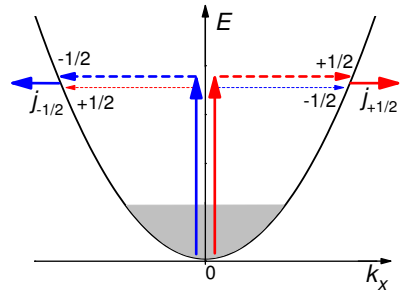}} \caption{Microscopic origin
of pure spin current induced under light absorption by free
electrons. The free-carrier absorption is a combined process
involving electron-photon interaction (vertical solid lines) and
electron scattering (dashed horizontal lines).} \vspace{-0.5cm}
\end{figure}

Let us assume the photon energy $\hbar\omega$ to exceed the
typical electron kinetic energy, $E_F$ for degenerate and $k_B T$
for non-degenerate electron gas. Then the pure spin current
induced by free-carrier light absorption is given by
\begin{equation}
F_{x}^{\alpha}=\frac{\tau_e}{\hbar} \left[ \frac{V_{\alpha
x}}{V_0} \left( 1+ \frac{|e_x|^2-|e_y|^2}{2} \right) +
\frac{V_{\alpha y}}{V_0}\ e_x e_y \right] \eta_{e1} I\ ,
\end{equation}
where $\bm{e}=(e_x,e_y)$ is the light polarization unit vector,
and $\eta_{e1}$ is the light absorbance in this spectral range.

In addition to the free-carrier absorption, the spin-de\-pen\-dent
asymmetry of electron-pho\-non interaction can also give rise to a
pure spin current in the process of photoelectron energy
relaxation. In this relaxational mechanism the spin current is
generated in a system of hot carriers, independently of heating
means.

Besides the spin, free charge carriers can be characterized by
another internal property, e.g., by the valley index in
multi-valley semiconductors. Thus, one can consider pure
orbit-valley currents in which case the net electric current
vanishes but the partial currents contributed by carriers in the
particular valleys are nonzero.

\subsection*{Acknowledgements} We acknowledge helpful discussions
with V.V.~Bel'kov and S.D.~Ganichev. This work was supported by
RFBR, INTAS, programs of the RAS and Foundation ``Dynasty'' -
ICFPM.

\vspace{0.2cm}

\noindent \textbf{References}

\noindent [1]~~R.D.R.~Bhat, F.~Nastos, A.~Najmaie, and
J.E.~Sipe,\\ \mbox{}~~~~~arXiv:cond-mat/0404066.

\noindent [2]~~E.L.~Ivchenko and S.A.~Tarasenko, Zh. Eksp. Teor.\\
\mbox{}~~~~~Fiz. {\bf 126}, 476 (2004) [JETP {\bf 99}, 379
(2004)].

\end{document}